\documentclass{article}
\usepackage[T1]{fontenc} 
\usepackage{graphicx} 
\usepackage{color} 
\usepackage{amsmath,amssymb,theorem} 
\usepackage{listings} 
\usepackage{booktabs} 
\usepackage{xspace} 
\usepackage[printonlyused,withpage]{acronym} 
\usepackage{rotating} 
\usepackage{microtype} 
\usepackage{braket}
\usepackage{bm}
\usepackage{mathabx}
\usepackage[algo2e]{algorithm2e}
\usepackage{float}
\usepackage{epstopdf}
\usepackage{authblk}

\newcommand{\up}{\uparrow}
\newcommand{\down}{\downarrow}

\begin{document}

\title{Maximum Probability Domains for Hubbard Models}

\author[1]{Guillaume Acke}
\author[1,2]{Stijn De Baerdemacker}
\author[2]{Pieter W. Claeys}
\author[1]{Mario Van Raemdonck}
\author[2]{Ward Poelmans}
\author[2]{Dimitri Van Neck}
\author[1]{Patrick Bultinck\thanks{Corresponding author. Email: Patrick.Bultinck@UGent.be}}
\affil[1]{Department of Inorganic and Physical Chemistry, Ghent University, 9000 Ghent, Belgium}
\affil[2]{Center for Molecular Modeling, Ghent University, 9052 Zwijnaarde, Belgium}
\date{}
  
\renewcommand\Authands{ and }

\maketitle

\begin{abstract}
  The theory of Maximum Probability Domains (MPDs) is formulated for the Hubbard model in terms of projection operators and generating functions for both exact eigenstates as well as Slater determinants. A fast MPD analysis procedure is proposed, which is subsequently used to analyse numerical results for the Hubbard model. It is shown that the essential physics behind the considered Hubbard models can be exposed using MPDs. Furthermore, the MPDs appear to be in line with what is expected from Valence Bond Theory-based knowledge.


\end{abstract}

\section{Introduction} \label{sec:introduction}

Chemists often rely on a set of concepts that are not directly derivable from quantum mechanical observables \cite{faraday_disc2007}. As such, these tools are not uniquely defined from first principles, and are highly prone to user bias. Nevertheless, despite the inherent arbitrariness behind them, these tools still constitute a major part of chemical theory as they have proven their value over and over again for decades.
\par
One of these tools is the theory of Lewis structures, which was proposed by Lewis in 1916 \cite{lewis1916a}, only 20 years after the discovery of the electron and 10 years before Schr\"{o}dinger's seminal paper \cite{schrodinger1926a}. The visionary character of Lewis can hardly be overestimated as the far majority of electronic structures, especially in organic chemistry, are based on Lewis's rules. The combination of Lewis structures along with Valence Shell Electron Pair Repulsion (VSEPR) theory \cite{nyholm1957a, gillespie1970a} also laid the foundations of structural chemistry.
\par
In view of the marvel of Lewis's theory, major problems remain for modern day theoretical chemists in explaining this success. Lewis structures rely on non-observable properties, just as, for instance, atoms-in-molecules (AIM) \cite{mulliken1955a, hirshfeld1977a, bader1991a, bultinck2009a} and the chemical bond. In this context, ``non-observable'' indicates that these concepts cannot be extracted uniquely from the wave function using a Hermitian operator based on the correspondence principle. This does not mean that these concepts are \textit{per se} orthogonal to quantum mechanics but it does mean that there exist multiple ways to extract them from a wave function, in the best case using well argued but nonetheless biased approaches.
\par
In the present work we discuss a method that does not introduce the concept of Lewis structures into quantum mechanics but rather lets them --- when applicable --- emanate from the structure of the wave function $\Psi$ without reference to an AIM model. This central idea sets this method apart from methods based on e.g. delocalisation indices \cite{ponec1994a, wiberg1968a, fradera1999a}, also known as shared electron distribution indices (SEDIs) \cite{ponec2005a}, where first an AIM method is chosen to identify two bonded atoms or atomic domains or more in the case of multicenter bonding \cite{giambiagi1990a, bultinck2005a}.
\par
According to the Copenhagen interpretation of quantum mechanics, a probability is obtained when the quantity $\left| \Psi \right|^2$ is multiplied by an infinitesimal volume element. For an $N$-electron wave function $\Psi(\mathbf{x}_1, \ldots, \mathbf{x}_N)$, where $\mathbf{x}$ denotes a space-spin coordinate $\mathbf{x} = (\mathbf{r}, \sigma)$ with $\mathbf{r} \in \mathbb{R}^3$ and $\sigma \in \{\uparrow, \downarrow\}$, the probability $p(1,\Omega)$ of finding an electron in a volume $\Omega \subset \mathbb{R}^3$ is given by summing over all $N$ spin-coordinates and integrating one electron over the domain $\Omega$ while integrating the remaining $N-1$ electrons over the entire space
\begin{equation}
p(1, \Omega) = N \sum\limits_{\sigma_1, \ldots, \sigma_N} \int\limits_{\Omega} \mathrm{d}\mathbf{r}_1 \underbrace{\int\limits_{\mathbb{R}^3} \mathrm{d}\mathbf{r}_2 \ldots \int\limits_{\mathbb{R}^3} \mathrm{d}\mathbf{r}_N}_{N-1} \left| \Psi(\mathbf{x}_1, \ldots, \mathbf{x}_N) \right|^2 \ .
\label{eq2}
\end{equation}
The prefactor $N$ corrects for the permutational symmetry of the $N$ electrons. A related, but different question is the probability $P(1, \Omega)$ of finding one (and only one) electron in a domain $\Omega$ with all other electrons outside this volume (or, equivalently, in the complementary domain $\overline{\Omega} = \mathbb{R}^3 \backslash \Omega$).
\begin{equation}
P(1, \Omega) = N \sum\limits_{\sigma_1, \ldots, \sigma_N} \int\limits_{\Omega} \mathrm{d}\mathbf{r}_1 \underbrace{\int\limits_{\overline{\Omega}} \mathrm{d}\mathbf{r}_2 \ldots \int\limits_{\overline{\Omega}} \mathrm{d}\mathbf{r}_N}_{N-1} \left| \Psi(\mathbf{x}_1, \ldots,\mathbf{x}_N) \right|^2 \ .
\label{eq3}
\end{equation}
This can be generalised to the probability $P(\nu, \Omega)$ of finding $\nu$ (and only $\nu$) electrons in $\Omega$, resulting in the probability
\begin{equation}
  P(\nu, \Omega) = \binom{N}{\nu} \sum\limits_{\sigma_1, \ldots, \sigma_N} \underbrace{\int\limits_{\Omega} \mathrm{d}\mathbf{r}_1 \ldots \int\limits_{\Omega} \mathrm{d}\mathbf{r}_{\nu}}_{\nu} \underbrace{\int\limits_{\overline{\Omega}} \mathrm{d}\mathbf{r}_{\nu+1} \ldots \int\limits_{\overline{\Omega}} \mathrm{d}\mathbf{r}_N}_{N-\nu}  \left| \Psi(\mathbf{x}_1,\mathbf{x}_2,\ldots,\mathbf{x}_N) \right|^2 \ .
\label{eq4}
\end{equation}
When the domain $\Omega$ is related to some AIM, these formulae link directly to Domain Averaged Fermi Holes \cite{ponec1997a}, SEDI \cite{ponec2005a} and multicenter indices \cite{giambiagi1990a, bultinck2005a, francisco2007a, pendas2007a, pendas2007b, pendas2007c, francisco2008a, francisco2014a}. Interestingly, Savin showed that chemical interpretations can also be associated to those domains whose shapes have been optimised so as to maximise these probabilities. The resulting domains are called the `Maximum Probability Domains' or MPDs \cite{savin2002a}. An $\mathrm{MPD}(\nu)$ is a domain $\Omega^{\ast}$ for which the probability of finding $\nu$ (and only $\nu$) electrons is maximal
\begin{equation}
  P(\nu,\Omega^{\ast}) = \max_{\Omega} P(\nu, \Omega) \ ,
  \label{eq:introduction:max}
\end{equation}
where the maximality criterion is understood to include local maxima. This makes the domain the result of a well-defined criterion without inferring it from an AIM model. The underlying hope is that the resulting MPDs will capture those regions in space where the bonding pairs of the Lewis model would appear. In 2004, Canc\`{e}s  et al. \cite{cances2004a} devised an algorithm to perform this optimisation, spurring renewed interest in MPDs \cite{chamorro2003a, gallegos2005a, scemama2005a, scemama2007a, causa2011a, causa2011b, causa2013a, menendez2015a, causa2015a}.
\par
Despite many interesting results from probabilities computed from user specified domains, only relatively little work has appeared where true MPD's are computed, i.e. where the domains are actually optimised under probability criteria. This is due to several bottlenecks. The optimisation procedure has currently only been derived for single determinant wave functions, which are known to fail for many interesting molecules where correlation is important. Furthermore, technical issues with the optimisation algorithms lead to numerical errors, which cloud the fundamental properties and understanding of MPDs. Such fundamental questions have only been addressed using analytical models \cite{lopes2012a, menendez2014a}.
\par
In the present work, we aim to go beyond purely analytical models by performing full configuration interaction (FCI) calculations for a discrete Hubbard model system. In this model system, both the domain probabilities as well as the optimisation of the domains can be formulated much more succinctly because the bottlenecks associated to optimising probability domains in $\mathbb{R}^3$ evaporate by discretising the space on a lattice. In quantum chemical terms, Hubbard models can be described as networks of hydrogens where a minimal basis set is used and a zero differential overlap occurs between the basis functions, rendering them orthonormal. The diagonalisation of the model Hamiltonian may be performed for different values of $U$ and $t$, known as the on-site and hopping terms in the Hubbard Hamiltonian \cite{hubbard1963a} (see Section \ref{sec:methodology}). The physics of the problem foretells what is to be expected for different limiting regimes and this model will thus serve well as a testing ground for MPD theory. For reasons explained below, MPDs are slightly generalised to a setup where a separate domain for $\nu_{\uparrow}$ (and only $\nu_{\uparrow}$) electrons with spin $\uparrow$ is sought as well as a possibly different domain for $\nu_{\downarrow}$ (and only $\nu_{\downarrow}$) electrons with spin $\downarrow$, for which the probabilities are given by:
\begin{multline}
  P(\nu_{\uparrow}, \uparrow, \Omega_{\uparrow}, \nu_{\downarrow}, \downarrow, \Omega_{\downarrow}) = \binom{N_{\uparrow}}{\nu_{\uparrow}} \binom{N_{\downarrow}}{\nu_{\downarrow}} \underbrace{\int\limits_{\Omega_{\uparrow}} \mathrm{d}\mathbf{r}_{1_{\uparrow}} \ldots \int\limits_{\Omega_{\uparrow}} \mathrm{d}\mathbf{r}_{\nu_{\uparrow}}}_{\nu_{\uparrow}} \underbrace{\int\limits_{\overline{\Omega}_{\uparrow}} \mathrm{d}\mathbf{r}_{\nu_{\uparrow}+1} \ldots \int\limits_{\overline{\Omega}_{\uparrow}} \mathrm{d}\mathbf{r}_{N_{\uparrow}}}_{N_{\uparrow}-\nu_{\uparrow}} \\
\underbrace{\int\limits_{\Omega_{\downarrow}} \mathrm{d}\mathbf{r}_{N_{\uparrow} + 1}  \ldots \int\limits_{\Omega_{\downarrow}} \mathrm{d}\mathbf{r}_{N_{\uparrow} + \nu_{\downarrow}}}_{\nu_{\downarrow}} \underbrace{\int\limits_{\overline{\Omega}_{\downarrow}} \mathrm{d}\mathbf{r}_{N_{\uparrow} + \nu_{\downarrow} + 1}  \ldots \int\limits_{\overline{\Omega}_{\downarrow}} \mathrm{d}\mathbf{r}_{N_{\uparrow} + N_{\downarrow}}}_{N_{\downarrow}-\nu_{\downarrow}} \\ |\Psi((\mathbf{r}_{1}, \uparrow), \ldots, (\mathbf{r}_{N_{\uparrow}}, \uparrow), (\mathbf{r}_{N_{\uparrow}+1},\downarrow),\ldots,(\mathbf{r}_{N_{\uparrow} + N_{\downarrow}}, \downarrow)|^2 \ ,
\label{eq6}
\end{multline}
with $N_{\uparrow}$ and $N_{\downarrow}$ the total number of spin-$\uparrow$ and spin-$\downarrow$ electrons respectively.
\par
The actual test case corresponds to a 6-site, 6-electron system with and without periodic boundary conditions, which chemically corresponds to a minimal Hubbard model of benzene and 1,3,5-hexatriene. Results of these calculations can be put in the context of the well-known work by H\"{u}ckel, whose theory very closely resembles Hubbard models \cite{huckel1930a}.

\section{Methodology} \label{sec:methodology}

\subsection{Hubbard models}
The most general Fermi-Hubbard model Hamiltonian allowing for intersite hopping between any $L$ sites, together with a site-dependent on-site repulsion is given by
\begin{equation}
  \hat{H}=-\sum_{i,j=1}^L\sum_{\sigma=\up,\down}t_{ij} \hat{a}_{i\sigma}^\dag \hat{a}_{j\sigma} +\sum_{i=1}^L U_i \hat{n}_{i\up}\hat{n}_{i\down} \ ,\label{theory:hamiltonian}
\end{equation}
with $t_{ij}$ an intersite-dependent spin-independent hopping, $U_i$ the site-dependent on-site repulsion depending on the local charges, and $\hat{a}_{i\sigma}^\dag$ and $\hat{a}_{i\sigma}$ the creation/annihilation operators of a particle with spin $\sigma$ ($\sigma=\up,\down$) on site $i$.  The number operator\footnote{We will explicitly use hat notation for the number operators $\hat{n}_{i\sigma}$ to distinguish between the operators and their eigenvalues $n_{i\sigma}$.} $\hat{n}_{i\sigma}=\hat{a}_{i\sigma}^\dag \hat{a}_{i\sigma}$ counts the number of particles with spin $\sigma$ on site $i$, with eigenvalues 0 and 1 because of the fermion anticommutation relations
\begin{equation}
  \{\hat{a}_{i\sigma},\hat{a}^\dag_{j\tau}\}=\delta_{ij}\delta_{\sigma\tau} \ ,\quad \{\hat{a}^\dag_{i\sigma},\hat{a}^\dag_{j\tau}\}=0 \ ,\quad \{\hat{a}_{i\sigma},\hat{a}_{j\tau}\}=0 \ .
\end{equation}
Although the adjacency matrix $t$ allows for hopping between any two sites in general, it usually follows the geometry of the lattice, which means that only hopping between neighbouring sites is considered.  In the same vein, the on-site repulsion is usually chosen site-independent because of symmetry considerations.  In the present paper (see Section \ref{sec:results_and_discussion}), we will consider a Hubbard model on a 1D chain with and without periodic boundary conditions, but it is clear that the following derivations are generally valid for Hubbard models.
\par
The eigenstates of the Hamiltonian can be expanded in the orthonormal basis spanned by the eigenstates of the number operators $\hat{n}_{i\sigma}$, 
\begin{equation}
  \ket{\Psi}=\sum_{[n]}c_{[n]}\ket{[n]} \ ,\label{theory:eigenstate}
\end{equation}
with $c_{[n]}$ the coefficient belonging to a basis state labelled by the partitioning $[n]=[[n_\up],[n_\down]]=[[n_{1\up},n_{2\up}\dots,n_{L\up}],[n_{1\down},n_{2\down},\dots,n_{L\down}]]$  of $n_\up$ spin-$\up$ and $n_\down$ spin-$\down$ particles over $L$ sites. Note that these basis states have in essence the same structure as the FCI basis vectors described by \cite{handy1980a} and can be represented accordingly as bitstrings \cite{olsen1988a}.  Both $n_\up$ and $n_\down$ are good quantum numbers for the eigenstates (\ref{theory:eigenstate}) due to the particle-number and spin-projection symmetry of the Hamiltonian (\ref{theory:hamiltonian}).  Explicitly, the basis states $\ket{[n]}$ can be expressed in Fock space
\begin{equation}\label{theory:sitebasis}
  \ket{[n]}=\bigotimes_{i=1}^L \ket{n_{i\up}} \bigotimes_{i=1}^L \ket{n_{i\down}} = \prod_{i=1}^L (\hat{a}_{i\up}^\dag)^{n_{i\up}}\prod_{i=1}^L (\hat{a}_{i\down}^\dag)^{n_{i\down}} \ket{\theta} \ ,
\end{equation}
with $\ket{\theta}$ the particle vacuum.  Using this representation, it is straightforward to verify that these states are eigenstates of the number operators $\hat{n}_{i\sigma}$
\begin{equation}
  \hat{n}_{i\sigma}\ket{[n]}=n_{i\sigma}\ket{[n]} \ ,
\end{equation}
leading to the interpretation of the number operator $\hat{n}_{i\sigma}$ as a projection operator for fermionic systems.  Indeed, if the site $i$ is not occupied with a spin-$\sigma$ particle ($n_{i\sigma}=0)$, then the state $\ket{[n]}$ is annihilated, whereas it is left unaltered otherwise ($n_{i\sigma}=1)$.  More general, the number operator $\hat{n}_{i\sigma}$ projects a general state onto that part of the Hilbert space that has the $i$th site occupied with a spin-$\sigma$ particle 
\begin{equation}
  \hat{n}_{i\sigma}\sum_{[n]}c_{[n]}\ket{[n]}=\sum_{[n]|n_{i\sigma}=1}c_{[n]}\ket{[n]} \ .
\end{equation}
Consequently, the \emph{probability} of finding a spin-$\sigma$ electron at site $i$ is also equivalent to the expectation value of the corresponding number operator
\begin{equation}\label{theory:expvalnumberoperator}
  \braket{\Psi|\hat{n}_{i\sigma}|\Psi}=\sum_{[n]|n_{i\sigma}=1}|c_{[n]}|^2 \ .
\end{equation}
Formally, the number operators $\hat{n}_{i\sigma}$ fulfil the requirements of projection operators thanks to the fermionic anticommutation relations
\begin{equation}
  \hat{n}_{i\sigma}^2=\hat{n}_{i\sigma} \ ,
\end{equation}
however they do not partition the Hilbert space into disjoint orthogonal subspaces due to 
\begin{equation}
  \hat{n}_{i\sigma}\hat{n}_{j\tau}\neq 0 \ , \quad \forall i,j,\sigma,\tau \ .
\end{equation}
The orthogonal complement of each projection operator $\hat{n}_{i\sigma}$ has a physical interpretation
\begin{equation}
  1-\hat{n}_{i\sigma}=1-\hat{a}^\dag_{i\sigma}\hat{a}_{i\sigma}=\hat{a}_{i\sigma}\hat{a}^\dag_{i\sigma} \ ,
\end{equation}
as the number operator counting the number of spin-$\sigma$ \emph{holes} at site $i$.  Like its complementary particle number operator, the hole number operator has the structure of a projection operator, so we will use the following definitions in the remainder of the paper for ease of notation
\begin{equation}
  \hat{O}_{i\sigma}:=\hat{n}_{i\sigma}\quad \hat{U}_{i\sigma}:=1-\hat{n}_{i\sigma} \ , \label{theory:projectionoperators}
\end{equation}
projecting against the orthogonal subspaces that are occupied ($O$) and unoccupied ($U$) with a spin-$\sigma$ respectively.  Note that $\hat{O}_{i\sigma}+\hat{U}_{i\sigma}=1$ and $\hat{O}_{i\sigma}\hat{U}_{i\sigma}=0$, $\forall i,\sigma$.
\subsection{MPDs using projection operators} \label{subsec:projection_operators}
In contrast to the intricate definition for continuous domains (see Section \ref{sec:introduction}), in discrete lattice models such as the Hubbard model, a \emph{domain} $\Omega$ can be defined as a subset of the total set of sites.  So, a domain $\Omega$ of size $M$ within a Hubbard model of size $L$ can also be represented as a partitioning 
\begin{equation}
\Omega=[\omega]=[\omega_1,\omega_2,\dots,\omega_L] \ ,
\end{equation}
with the restriction that $\omega_i=0$ or $1$ ($\forall i$), and $\sum_{i=1}^L$
$\omega_i=M$.  A domain $\Omega$ does not necessarily have to be connected, so there are $\binom{L}{M}$ different possibilities to pick a domain of size $M$, adding up to $2^L$ different domains of variable size in total.  Often it is more convenient to use a different notation for $\Omega$, as the subset $\{i_1, i_2, \ldots, i_M\}$ of the $M$ sites constituting the domain, so for which $\omega_i=1$.  For instance, a domain $\Omega$ of size $M=2$, consisting of the 1\textsuperscript{st} and 3\textsuperscript{rd} site  within a total of $L=5$ sites can be represented in two different ways: $\Omega=[1,0,1,0,0]=\{1,3\}$.
\par
Although the present definition of MPDs (Eq. \ref{eq:introduction:max}) is blind to the spin degree of freedom in the model, it is possible to resolve the spin dependency.  Not to complicate matters, we first focus on same-spin particles. The probability $P(\nu_\sigma,\sigma,\Omega)$ of finding $\nu_\sigma$ (and only $\nu_\sigma$) spin-$\sigma$ particles within a domain $\Omega$ can be conveniently formulated using the projection operators (\ref{theory:projectionoperators}).  Before introducing the general formula, it is instructive to discuss some special cases.  As pointed out in the previous section, the probability $P(1,\sigma,\{i\})$ of finding one (and only 1) spin-$\sigma$ particle at a single site $i$ (or domain $\Omega=\{i\}$ of size $M=1$) is simply the expectation value of the number of spin-$\sigma$ particles at that site, or equivalently, the expectation value of the projection operator
\begin{equation}
  P(1,\sigma,\{i\}) = \braket{\Psi|\hat{n}_{i\sigma}|\Psi} = \braket{\Psi|\hat{O}_{i\sigma}|\Psi} \ .
\end{equation}
Equivalent, the probability of finding $\nu_\sigma=0$ particles within $\Omega=\{i\}$ is
\begin{equation}
  P(0,\sigma,\{i\}) = \braket{\Psi|1-\hat{n}_{i\sigma}|\Psi} = \braket{\Psi|\hat{U}_{i\sigma}|\Psi} \ .
\end{equation}
The situation is a little more involved for the $M=2$ case.  The  probability of finding exactly $\nu_\sigma=0,1$, or $2$ can be written respectively as
\begin{align}
  P(0,\sigma,\{i_1,i_2\}) &= \braket{\Psi|\hat{U}_{i_1\sigma}\hat{U}_{i_2\sigma}|\Psi} \ , \\
  P(1,\sigma,\{i_1,i_2\}) &= \braket{\Psi|(\hat{O}_{i_1\sigma}\hat{U}_{i_2\sigma}+\hat{U}_{i_1\sigma}\hat{O}_{i_2\sigma})|\Psi} \ , \label{theory:mpd:1in2}\\
  P(2,\sigma,\{i_1,i_2\}) &= \braket{\Psi|\hat{O}_{i_1\sigma}\hat{O}_{i_2\sigma}|\Psi} \ .
\end{align}
Indeed, the projection operators $\hat{O}_{i\sigma}$ and $\hat{U}_{i\sigma}$ make sure that only those components $\ket{[n]}$ of the state $\ket{\Psi}$ survive that correspond to $\nu_\sigma=1$ within the domain $\{i_1,i_2\}$.  The combined use of $\hat{O}$ and $\hat{U}$ projection operators in $P(1,\{i_1,i_2\})$ is essential to ascertain that one (and only one) particle is contained within the domain. 
From these examples, the general formulation of the probabilities in terms of projection operators comes naturally.  Explicitly, the probability to find $\nu_\sigma$ (and only $\nu_\sigma$) particles within a domain $\Omega=\{i_1, \ldots, i_M\}$ is given by
\begin{equation}
  P(\nu_\sigma,\sigma,\Omega) = \braket{\Psi|\hat{P}(\nu_\sigma,\sigma,\Omega)|\Psi} \ ,
\end{equation}
with the probability domain operator defined as 
\begin{equation}\label{theory:mpd:operator}
\hat{P}(\nu_\sigma,\sigma,\Omega) = \frac{1}{\nu_\sigma!(M-\nu_\sigma)!}\sum_{\pi\in S_M}\prod_{k=1}^{\nu_\sigma}\hat{O}_{i_{\pi(k)}\sigma}\prod_{k=\nu_\sigma+1}^M\hat{U}_{i_{\pi(k)}\sigma} \ ,
\end{equation}
and $\pi$ a permutation of the $M$ sites within the domain $\Omega$ ($\pi$ is an element of $S_M$, the permutation group over $M$ elements).  The prefactor corrects for double counting.  Because the projection operators are consistent with the interpretation as probabilities, we will omit the expectation value notation, and continue with the projection operator formulation (\ref{theory:mpd:operator}).

\subsection{Generating function}

\subsubsection{General case}

The generating function operator
\begin{equation}
  \hat{G}(\sigma,\Omega,t)=\sum_{\nu_\sigma=0}^{M}\hat{P}(\nu_\sigma,\sigma,\Omega)t^{\nu_\sigma} \ ,
\end{equation}
condenses all possible probabilities for a given domain $\Omega$ into one polynomial function in $t$ of degree $M$.  It is straightforward to prove (see Appendix \ref{subsection:appendix:factorisation}) that the generating function operator for a domain $\Omega=\{i_1, \ldots, i_M\}$ of size $M$ can be factorised as 
\begin{equation}\label{theory:mpd:genfunction}
  \hat{G}(\sigma,\Omega,t)=\prod_{k=1}^M(\hat{U}_{i_k\sigma}+t\hat{O}_{i_k\sigma}) \ ,
\end{equation}
or, equivalently
\begin{equation}\label{theory:generatingfunction:exponential}
  \hat{G}(\sigma,\Omega,t)=t^{\hat{n}_{\sigma}(\Omega)} \ ,
\end{equation}
in which $\hat{n}_\sigma(\Omega)$ counts the number of sites in the domain $\Omega$ that are occupied by a spin-$\sigma$ particle.  When evaluating the generating function operator for a quantum state $\ket{\Psi}$, it is convenient to express $\ket{\Psi}$ in the site basis spanned by the basis states (\ref{theory:sitebasis}).  We obtain
\begin{align}
  G(\sigma,\Omega,t) &=  \braket{\Psi|\hat{G}(\sigma,\Omega,t)|\Psi} \\
  &=\sum_{[n],[m]}c^\ast_{[n]}c_{[m]} \braket{[n]|\hat{G}(\sigma,\Omega,t)|[m]} \ .
\end{align}
All projection operators (\ref{theory:projectionoperators}) are diagonal in the site basis, so the generating function reduces to diagonal elements only,
\begin{equation}
  G(\sigma,\Omega,t)=\sum_{[n]}|c_{[n]}|^2 \braket{[n]|\hat{G}(\sigma,\Omega,t)|[n]} \ .
\end{equation}
The factorised expression for the generating function operator (\ref{theory:mpd:genfunction}) now comes in handy, because one can evaluate each factor separately.  Making use of the explicit notation for $\ket{[n]}$ in Eq.\ (\ref{theory:sitebasis}), one obtains
\begin{equation}
\braket{ [n]|\hat{G}(\sigma,\Omega,t)|[n]} = \prod_{k=1}^M \braket{ n_{i_k\sigma}|\hat{U}_{i_k\sigma}+t\hat{O}_{i_k\sigma}|n_{i_k\sigma}} = \prod_{k=1}^M(1-n_{i_k\sigma}+tn_{i_k\sigma}) \ ,
\end{equation}
in which we have used the diagonal property of the projector operators in the site basis.  Because $n_{i_k\sigma}$ is either 1 (the site $i_k$ in $\Omega$ is occupied with a spin-$\sigma$ particle in state $\ket{[n]}$) or 0 (the site $i_k$ in $\Omega$ is \emph{not} occupied with a spin-$\sigma$ particle in state $\ket{[n]}$), the formula reduces to
\begin{equation}
  \braket{[n]|\hat{G}(\sigma,\Omega,t)|[n]}=t^{n_\sigma(\Omega)} \ ,
\end{equation}
in which $n_\sigma(\Omega)$ is the number of sites in the domain $\Omega$ that are occupied by a spin-$\sigma$ particle in the state $\ket{[n]}$, consistent with the exponential form of the generating function operator (\ref{theory:generatingfunction:exponential}).   For a general state, one gets the result
\begin{equation}\label{theory:mpd:gensum}
  G(\sigma,\Omega,t)=\sum_{[n]}|c_{[n]}|^2t^{n_\sigma(\Omega)} \ .
\end{equation}
It is now possible to extract the probabilities $P(\nu_\sigma,\sigma,\Omega)$ by filtering those coefficients that have $\nu_\sigma=n_{\sigma}(\Omega)$ particles in the domain $\Omega$
\begin{equation}
P(\nu_\sigma,\sigma,\Omega)=\sum_{[n]}|c_{[n]}|^2\delta_{\nu_\sigma,n_{\sigma}(\Omega)} \ .
\end{equation}
\subsubsection{Slater determinants}
The factorisation of the generating function (\ref{theory:mpd:genfunction}) holds for arbitrary states and is not necessary limited to eigenstates of a Hamiltonian. However, as can be seen from (\ref{theory:mpd:gensum}), an exact evaluation of the generating function requires a summation over the whole Hilbert space, limiting the practical applicability to small systems.  This problem was previously encountered for MPDs in continuous space.  Nevertheless, it was shown how the generating function and resulting probabilities can be efficiently evaluated for a Slater determinant \cite{cances2004a}. These results can be immediately generalised towards lattice models. Taking $\ket{\Phi}$ as a Slater determinant with $N=n_{\uparrow}+n_{\downarrow}$ orthonormal occupied states, the generating function can be shown to reduce to
\begin{equation}\label{theory:mpd:slater1}
  \braket{\Phi | \prod_{k=1}^M(\hat{U}_{i_k\sigma}+t\hat{O}_{i_k\sigma}) | \Phi} = \det \left\{ \left[ I_N-S_\sigma(\Omega) \right]+t S_\sigma(\Omega) \right\} \ ,
\end{equation}
with $I_N$ the $N \times N$ identity matrix and $S_\sigma(\Omega)$ the $N \times N$ symmetric matrix defined as
\begin{equation}
  S_\sigma(\Omega)_{\alpha \beta}= \braket{\alpha|\sum_{k=1}^M \hat{n}_{i_k\sigma} |\beta} = \braket{\alpha|\hat{n}_{\sigma}(\Omega)|\beta} \ ,
\end{equation}
where $\alpha, \beta=1 \dots N$ label the occupied single-particle states in the Slater determinant. A dual expression can be found as
\begin{equation}\label{theory:mpd:slater2}
 \braket{ \Phi | \prod_{k=1}^M(\hat{U}_{i_k\sigma}+t\hat{O}_{i_k\sigma}) | \Phi } = \det \left\{ \left[ I_M-T_\sigma(\Omega) \right]+t T_\sigma(\Omega) \right\} \ ,
\end{equation}
with $I_M$ the $M \times M$ identity matrix and $T_\sigma(\Omega)$ the $M \times M$ symmetric matrix defined as
\begin{equation}
T_\sigma(\Omega)_{ij}= \braket{i\sigma|\sum_{\alpha =1}^N \hat{n}_{\alpha}|j\sigma} \ ,
\end{equation}
where $\ket{i\sigma}=\hat{a}^{\dag}_{i\sigma}\ket{\theta}$, with $i,j$ occupied sites in $\Omega$, and $\hat{n}_{\alpha}$ is the occupation number of the orbital $\alpha$. Starting from the factorised expression for the generating function, the proof for these identities is analogous to the one presented previously \cite{cances2004a} and is given in Appendix \ref{subsection:appendix:slater}. 

The size of these matrices scales with the number of occupied orbitals and the size of $\Omega$ respectively, and both the matrix elements and the determinant can be calculated in a polynomial time. Furthermore, all probabilities $P(\nu_{\sigma},\sigma,\Omega)$ can be expressed in the eigenvalues of these matrices \cite{cances2004a}. However, it should be stressed that these results present a trade-off: starting from a Slater determinant allows for an efficient calculation of the MPDs of an approximate state, whereas starting from the exact solution allows for an involved calculation of the MPDs of the exact state. In the present paper, we will limit ourselves to calculations using the exact ground state.
\subsection{Spin dependency of the MPD}
The extra spin degree of freedom complicates matters because a single site $i$ within a domain $\Omega$ can now be occupied with a spin-$\up$ particle, a spin-$\down$ particle, or both.  Consequently, the site number operator $\hat{n_i}=\hat{n}_{i\up}+\hat{n}_{i\down}$ loses its interpretation as a projection operator
\begin{equation}
\hat{n}_i^2=(\hat{n}_{i\up}+\hat{n}_{i\down})^2=\hat{n}_{i\up}+\hat{n}_{i\down}+2\hat{n}_{i\up}\hat{n}_{i\down}\neq \hat{n}_i \ ,
\end{equation}
so the probabilities need to be broken down into its spin projections $\sigma=\up,\down$.  At its most generality, the probability of finding $\nu_\up$ (and only $\nu_\up$) and $\nu_\down$ (and only $\nu_\down$) in respective domains $\Omega_\up$ and $\Omega_\down$ is given at the operator level
\begin{equation}\label{theory:spin:combined}
\hat{P}(\nu_\up,\Omega_{\up};\nu_\down,\Omega_\down)=\hat{P}(\nu_\up,\up,\Omega_{\up})\hat{P}(\nu_\down,\down,\Omega_\down) \ .
\end{equation}
Note that the domains $\Omega_\up$ and $\Omega_\down$ do not need not to be taken equal in general, however it is often in the interest of interpretation to do so ($\Omega_\up=\Omega_\down=\Omega$).  The differentiation between the two spin degrees of freedom necessitates the introduction of an additional degree of freedom in the generating function operator
\begin{equation}\label{theory:spin:generatingfunction}
\hat{G}(\Omega_\up,t;\Omega_\down,s)=\sum_{\nu_\up=0}^{M_\up}\sum_{\nu_\down=0}^{M_\down}\hat{P}(\nu_\up,\Omega_{\up};\nu_\down,\Omega_\down)t^{\nu_\up} s^{\nu_\down} \ .
\end{equation}
The probabilities are nicely factorised (\ref{theory:spin:combined}), which easily extends to the generating function
\begin{align}
\hat{G}(\Omega_\up,t;\Omega_\down,s)&=\sum_{\nu_\up=0}^{M_\up}\hat{P}(\nu_\up,\up,\Omega_\up)t^{\nu_\up}\sum_{\nu_\down=0}^{M_\down}\hat{P}(\nu_\down,\down,\Omega_\down) s^{\nu_\down}\\
&=\hat{G}(\up,\Omega_\up,t)\hat{G}(\down,\Omega_\down,s) \ ,
\end{align}
implying that many properties of the generating function (such as the exponential form (\ref{theory:generatingfunction:exponential}), or the expectation value for a single Slater determinant) can be simply transferred from the single spin-$\sigma$ case.

Multiple ``spin-averaged'' properties can now be extracted from the spin-resolved generating function (\ref{theory:spin:generatingfunction}).  For ease of interpretation, we will choose $\Omega_\up=\Omega_\down=\Omega$, but everything can be formulated with distinct domains
\begin{itemize}
\item The probability of finding $\nu$ particles (\emph{in}dependent of the spin, so $\nu=\nu_\up+\nu_\down$) in a domain $\Omega$ can be found by putting $s=t$ in the generating function, and reorganising the summation
\begin{equation}
\hat{G}(\Omega,t;\Omega,t)= \sum_{\nu=0}^{2M}\left(\sum_{\nu_\up=\max(\nu-M,0)}^{\min(M,\nu)}\hat{P}(\nu_\up,\up,\Omega)\hat{P}(\nu-\nu_\up,\down,\Omega)\right)t^{\nu} \ .
\end{equation}
The probability of finding $\nu$ (and only $\nu$) particles is obtained by taking the coefficient of $t^\nu$ (assume $\nu$ below or equal to half filling of the domain $\nu\le M$, the other case is symmetric)
\begin{equation}
\hat{P}_{\up\down}(\nu,\Omega):=\sum_{\nu_\up=0}^{\nu}\hat{P}(\nu_\up,\up,\Omega)\hat{P}(\nu-\nu_\up,\down,\Omega) \ .
\end{equation} 
Note that this definition of $\hat{P}_{\up\down}(\nu,\Omega)$ coincides with the original definition of MPDs, \emph{in}dependent of the spin of the particles that are in the domain. 
\item An interesting variant of the previous definition is the probability of finding $\nu_\up$ spin-$\up$ particles in a domain $\Omega$, irrespective of how many spin-$\down$ particles there are.  This can be found by substituting $s=1$ in the generating function (\ref{theory:spin:generatingfunction})
\begin{equation}
\hat{G}(\Omega,t;\Omega,1)=\hat{G}(\up,\Omega_\up,t)\hat{G}(\down,\Omega_\down,1)=\hat{G}(\up,\Omega_\up,t) \ .
\end{equation}
The probability of finding $\nu_\up$ spin-$\up$ particles then coincides with the definition of the spin-dependent probabilities defined previously (\ref{theory:mpd:operator})
\begin{equation}
\hat{P}_\up(\nu_\up,\Omega):=\hat{P}(\nu_\up,\up,\Omega) \ .
\end{equation}
The same argument holds for the spin-$\down$ probabilities when taking $t=1$\ .

\end{itemize}  

\section{Implementation}

For the calculation of the MPDs, we assume that an eigenvector $\ket{\Psi}$ of the relevant Hubbard model Hamiltonian $\hat{H}$ has been determined and can be expressed in the site basis $\left\{ \ket{[n]} \right\}$ with coefficients $\left\{ c_{[n]} \right\}$. If we constrain all spin-resolved domains $\Omega_{\uparrow}, \Omega_{\downarrow}$ to occupy the same sites, such that for all spin-unresolved domains $\Omega$ the following is valid: $\forall \Omega: \Omega_{\uparrow} = \Omega_{\downarrow} = \Omega$, then we can construct $2^L$ domains in total, with $L$ the number of sites. Since we can represent such domains $\Omega \in \left\{ \Omega \right\}, \left| \{\Omega\} \right| = 2^L$  in the same site basis $\left\{ \ket{[n]} \right\}$, we can store them as bitstrings (Section \ref{subsec:projection_operators}). As such, we can determine the associated probabilities as follows:

\begin{algorithm2e}
  \KwData{$\left\{ \ket{[n]} \right\}$ (explicit or via addressing scheme), $\left\{c_{[n]}\right\}, \left\{ \Omega \right\}$}
  \KwResult{$P(\nu, \Omega) (\forall \Omega \in \left\{\Omega \right\})$ }
  \For{$\ket{[n]}$ in $\left\{ \ket{[n]} \right\}$} {
    \For{$\Omega$ in $\left\{ \Omega \right\}$} {
      \If{POPCNT $(\ket{[n]} \land \Omega) == \nu$} {
        $P(\nu, \Omega) \mathrel{+}= \left| c_{[n]} \right|^2$
      }
    }
  }
\end{algorithm2e}

In this algorithm, POPCNT counts the number of `1' bits in the passed bitstring (this functionality is available as an intrinsic procedure in most Fortran and C/C++ compilers).
\par
Since the definition of MPDs (Eq. (\ref{eq:introduction:max})) allows for local maxima, we can define the set of MPDs for $\nu$ electrons in the present context as follows:
\begin{equation}
  \mathrm{MPDs}(\nu) = \left\{ \Omega^{*} | \forall \Omega \in \epsilon_{\Omega^{*}}: P(\nu, \Omega^{*}) \geq P(\nu, \Omega)  \right\} \ ,
\label{eq5}
\end{equation}
where $\epsilon_{\Omega^{*}}$ defines the set of domains which are within a certain `distance' $\epsilon$ from the domain $\Omega^{*}$
\begin{equation}
  \epsilon_{\Omega^{*}} = \left\{ \Omega | \Vert \Omega^{*} - \Omega \Vert \leq \epsilon \right\} \ .
\end{equation}
In this work, we use the Hamming distance as a distance measure \cite{hamming1950a}. If we set $\epsilon = 1$, this reduces to a single-bit flip stability criterion. Only domains which are maximal in their probability within this distance are retained as MPDs.
\par
If we suppose that the computational cost for diagonalising the Hubbard-model for $L$ sites is of the order of the dimension of the Hilbert space $\mathcal{H}$ cubed,
\begin{equation}
  C_{\mathrm{Hubbard}} = \left[ \mathrm{dim} \mathcal{H} \right]^3 \ ,
\end{equation}
and the cost for the MPD procedure is of the order of the product of the number of coefficients, the number of domains and the number of single-bit flips for a given domain,
\begin{equation}
  C_{\mathrm{MPD}} = \left[\mathrm{dim} \mathcal{H} \right] 2^L (L-1) \ ,
\end{equation}
then for Hubbard models at half-filling, imposing $S_z$ symmetry, $C_{\mathrm{Hubbard}} > C_{\mathrm{MPD}}$. Hence, for Hubbard models at half-filling, this analysis procedure will always take less time than the determination of the eigenvector to be analysed. This allows for the use of MPDs as an efficient interpretational tool, since they can be calculated at a fraction of the computational cost of the determination of the eigenvector to be analysed.

\section{Applications: 1D Hubbard with 6 sites at half-filling} \label{sec:results_and_discussion}

As discussed in Section \ref{sec:introduction}, we analyse the 1D Hubbard model with 6 sites at half filling (denoted as $6s,6e$ in the accompanying plots), with and without periodic boundary conditions, for a $U/t$ range from 0 to 100 in unit steps. For every value of $U/t$, the corresponding Hamiltonian is diagonalised through a FCI calculation and the resulting ground-state wave function is used as input for the domain probability computation. The domains are then tested for stability against single-bit flips, resulting in a set of (point group) symmetric classes of MPDs. In all cases, only one representative structure of a symmetry equivalent class of MPDs will be plotted.
\par
In the following we will focus on the set of MPDs for two electrons regardless of their spin (i.e. $\left\{\mathrm{MPD}(\nu)\right\} = \left\{ \Omega^{*} | P(\nu,\Omega^{*}) = \max_{\Omega} P(\nu, \Omega) \right\}$ with $\nu = 2$) as well as the set of MPDs for spin-resolved electrons (i.e. $\left\{\mathrm{MPD}(\nu_{\uparrow}, \nu_{\downarrow}) \right\} = \left\{\Omega^{*} | P(\nu_{\uparrow}, \uparrow, \nu_{\downarrow}, \downarrow, \Omega^{*}) = \max_{\Omega} P(\nu_{\uparrow}, \uparrow, \nu_{\downarrow}, \downarrow, \Omega) \right\}$ with $\nu_{\uparrow} + \nu_{\downarrow} = 2)$.

\subsection{With periodic boundary conditions: Hubbard benzene}

\begin{figure}[p]
  \begin{center}
    \includegraphics[width=\textwidth]{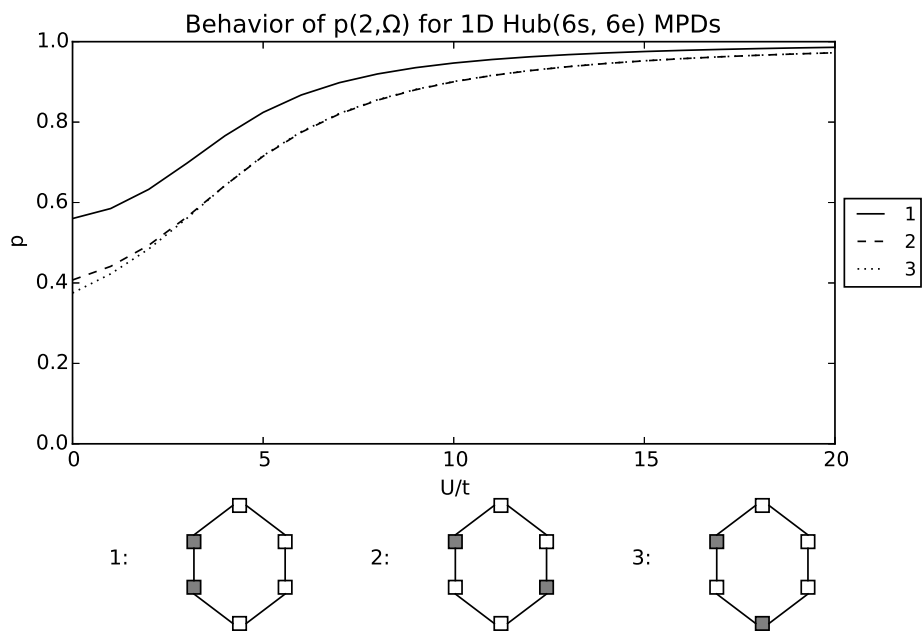}
    \caption{MPDs for choosing 2 spin-unresolved electrons for 1D Hubbard with periodic boundary conditions for 6 sites at half-filling for different $U/t$ values.}
    \label{fig:hub1_6s_6e_per_mpd_2e}      
  \end{center}
\end{figure}

The 1D Hubbard model with 6 sites at half filling and periodic boundary conditions can be used as a minimal model for benzene (Hubbard benzene). As shown in Figure \ref{fig:hub1_6s_6e_per_mpd_2e}, $\left\{\mathrm{MPD}(2)\right\}$ always consists of the same three classes of MPDs, irrespective of the $U/t$ value. This behaviour is not due to any constraints imposed, but stems from the structure of the wave function itself.
\par
In the large $U/t$ regime, the probabilities for each of the three two-site MPD classes become equal. At lower $U/t$ values, the probabilities of the classes differ. This is in line with what we expect from the underlying physics: in the large $U/t$ regime, we expect static correlation to become very important, in other words, we expect large on-site repulsion to force one electron per site, leading to the anti-ferromagnetic character of the wave function. At lower $U/t$ values, a mean field solution of the Hamiltonian dominates the FCI expansion. This solution should render the probabilities of the classes different.
\par
\begin{figure}[p]
  \begin{center}
  \includegraphics[width=\textwidth]{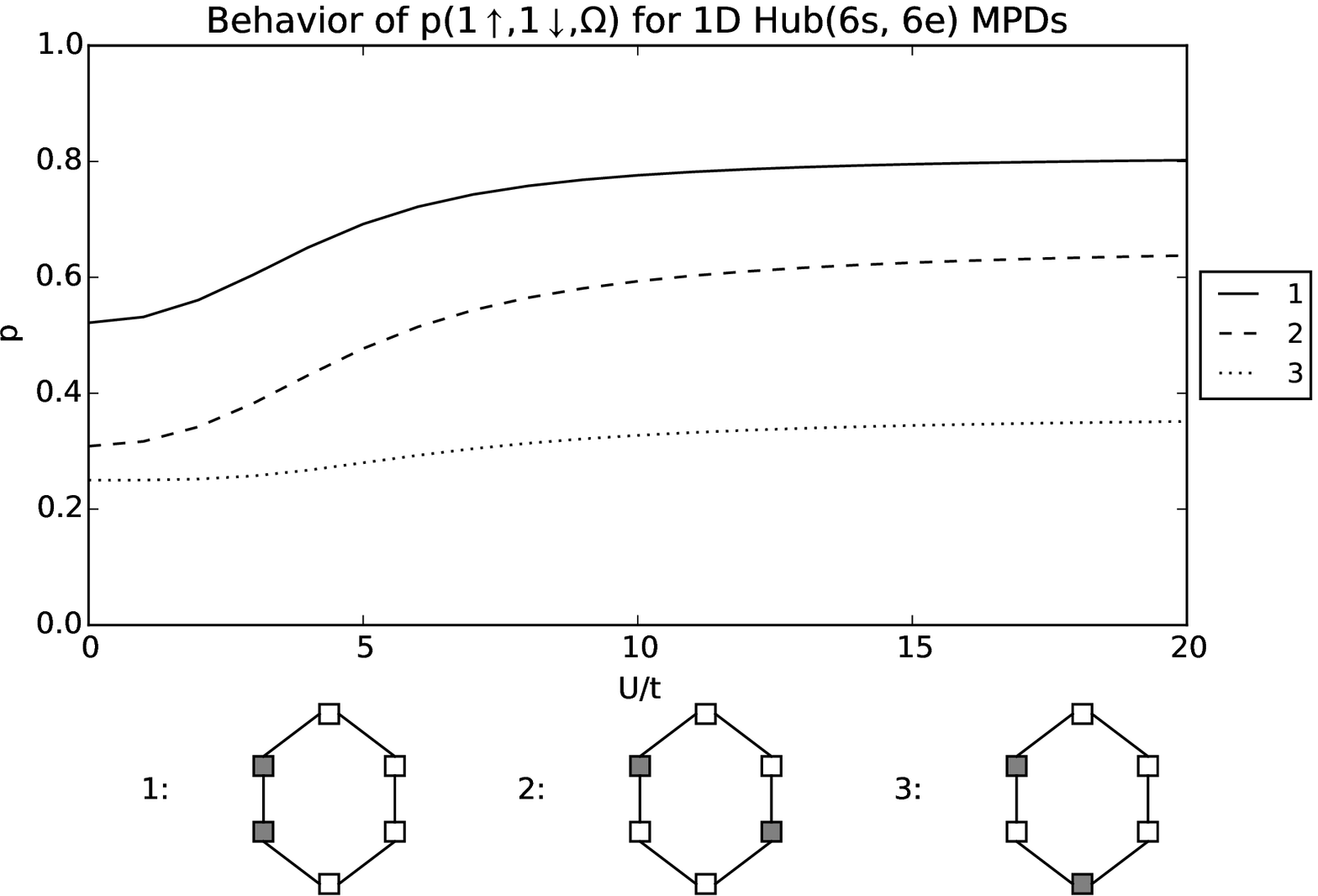}
  \caption{MPDs for choosing 1 up-spin and 1 down-spin electron for 1D Hubbard with periodic boundary conditions for 6 sites at half-filling for different $U/t$ values.}
  \label{fig:hub1_6s_6e_per_mpd_1u_1d}
  \end{center}
\end{figure}
\begin{figure}[p]
  \begin{center}
    \includegraphics[width=\textwidth]{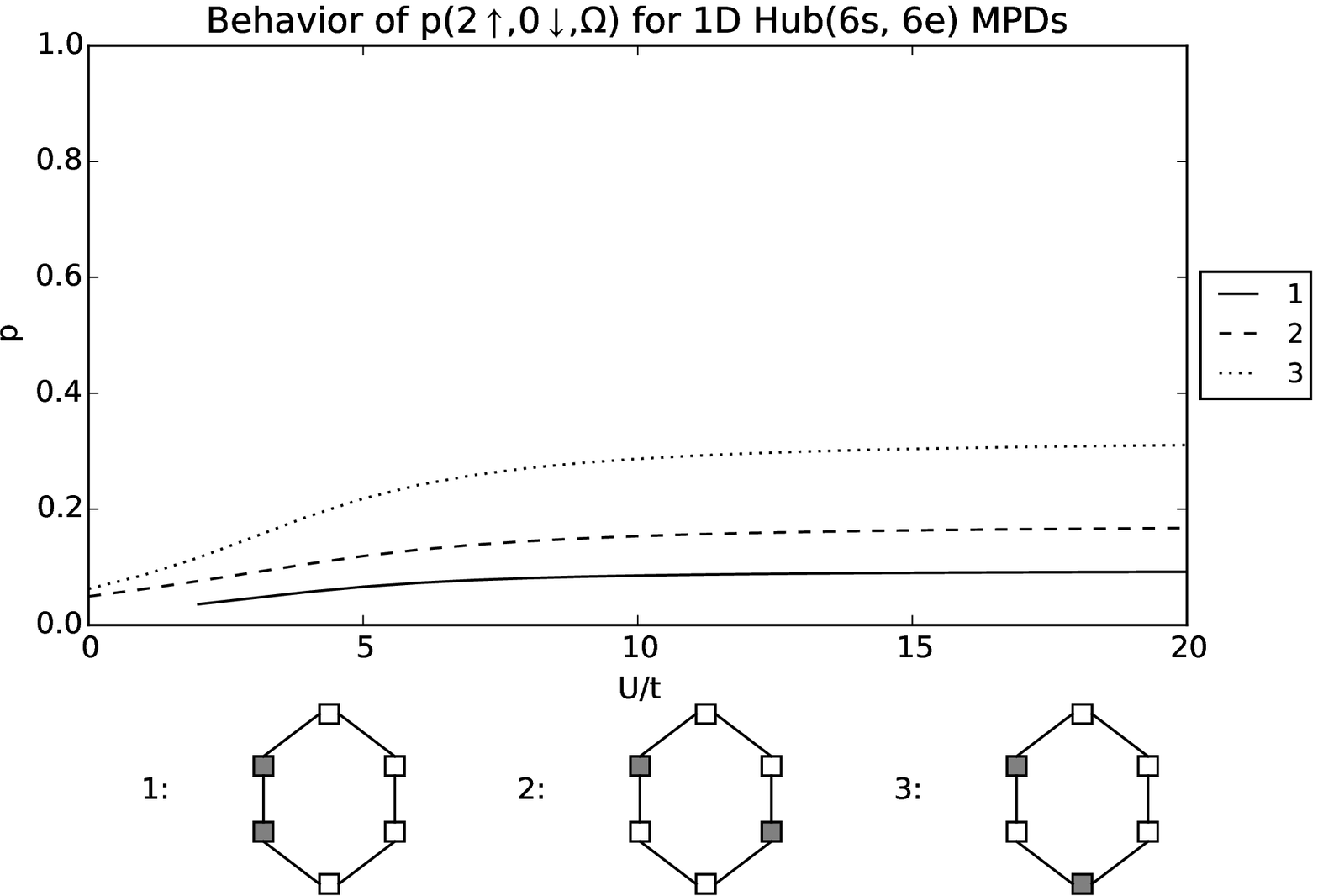}
    \caption{MPDs for choosing 2 up-spin and 0 down-spin electrons for 1D Hubbard with periodic boundary conditions for 6 sites at half-filling for different $U/t$ values.}
    \label{fig:hub1_6s_6e_per_mpd_2u_0d}    
  \end{center}
\end{figure}

When turning to spin-resolved MPDs we note that the theory itself does not require that $\left\{\mathrm{MPD}(2)\right\} = \left\{\mathrm{MPD}(1_{\uparrow}, 1_{\downarrow}) \right\} = \left\{\mathrm{MPD}(2_{\uparrow}, 0_{\downarrow}) \right\}$, nor that the ranking of the associated probability values should remain the same between groups. The coincidence that for benzene and for all $U/t$ values the same MPDs do appear is noteworthy (Figures \ref{fig:hub1_6s_6e_per_mpd_1u_1d} and \ref{fig:hub1_6s_6e_per_mpd_2u_0d}). This allows us to expand the spin-unresolved MPDs in terms of their spin-resolved counterparts. If the sets of MPDs would not have been the same, one could still calculate the probabilities associated with the non-maximal domains, but they should not be interpreted as maximum probability domains.
\par
Although the classes of MPDs remain the same, we can observe that in general the ranking of the probability values between the three groups does not remain the same. Figure \ref{fig:hub1_6s_6e_per_mpd_1u_1d} shows that MPD structure `1' remains the most probable when choosing 1 up-spin and 1 down-spin electron. This indicates that, for a typical electron pair consisting of one spin-up and one spin-down electron, the highest probability is found for an MPD containing two adjacent sites. However, this situation is reversed when choosing 2 up-spin and 0 down-spin electrons, where structure `3' is the most probable (Figure \ref{fig:hub1_6s_6e_per_mpd_2u_0d}).  Same-spin electrons therefore tend to remain in non-adjacent sites. In both spin-configurations, the second highest probability is associated with what could be called the para-delocalisation of class `2'.
\par
Note that the associated probabilities for the set of MPDs are quite small when choosing 2 up-spin and 0 down-spin electrons compared to the other two choices. Also, structure `1' is not stable against the single-bit flip criterion in the low $U/t$ regime, where a higher probability can be obtained by adding another site to the domain. However, the resulting three-site domain is again `unstable' and can flip to a genuine MPD, which is why three-site MPDs do not appear.
\par
\begin{figure}[p]
  \begin{center}
  \includegraphics[width=0.5\textwidth]{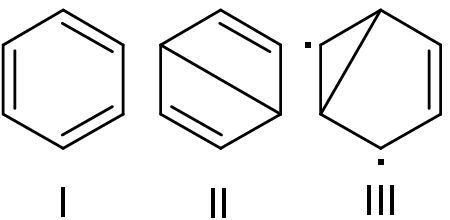}
  \caption{K\'{e}kule, Dewar and bi-radical structures of Hubbard benzene.}
  \label{fig:structures}    
  \end{center}
\end{figure}

It is tempting to relate these MPD findings to the well-known resonance structures of benzene, where the two Kekul\'{e} structures are the most important ones followed by the three Dewar structures shown as structure classes I and II respectively in Figure \ref{fig:structures}. The third type of MPD could be interpreted as related to class III in the same figure. It is indeed gratifying that, as was the case for delocalisation indices \cite{bader1996b, bader1974a}, the MPD results also seem to be in line with Valence Bond Theory results. However, our FCI wave functions are expressed in terms of orthogonal orbitals and as such, the individual valence bond wave functions for the different structures do not appear immediately in the FCI expansion. In future work, a CASVB \cite{thorsteinsson1996a, thorsteinsson1996b} type approach will be used to re-express --- without altering the energy --- the FCI wave function in terms of VB structures using a non-unitary transformation. As this lies outside the scope of the present work the apparent agreement momentarily remains to be treated with caution. However, from a global perspective, the MPDs do behave as expected for the different correlation regimes, from the mean field regime at zero $U$ to the high static correlation regime, and appear to be in line with what is expected from Valence Bond Theory-based knowledge.


\subsection{Without periodic boundary conditions: Hubbard 1,3,5-hexatriene}

\begin{figure}[p]
  \begin{center}
    \includegraphics[width=\textwidth]{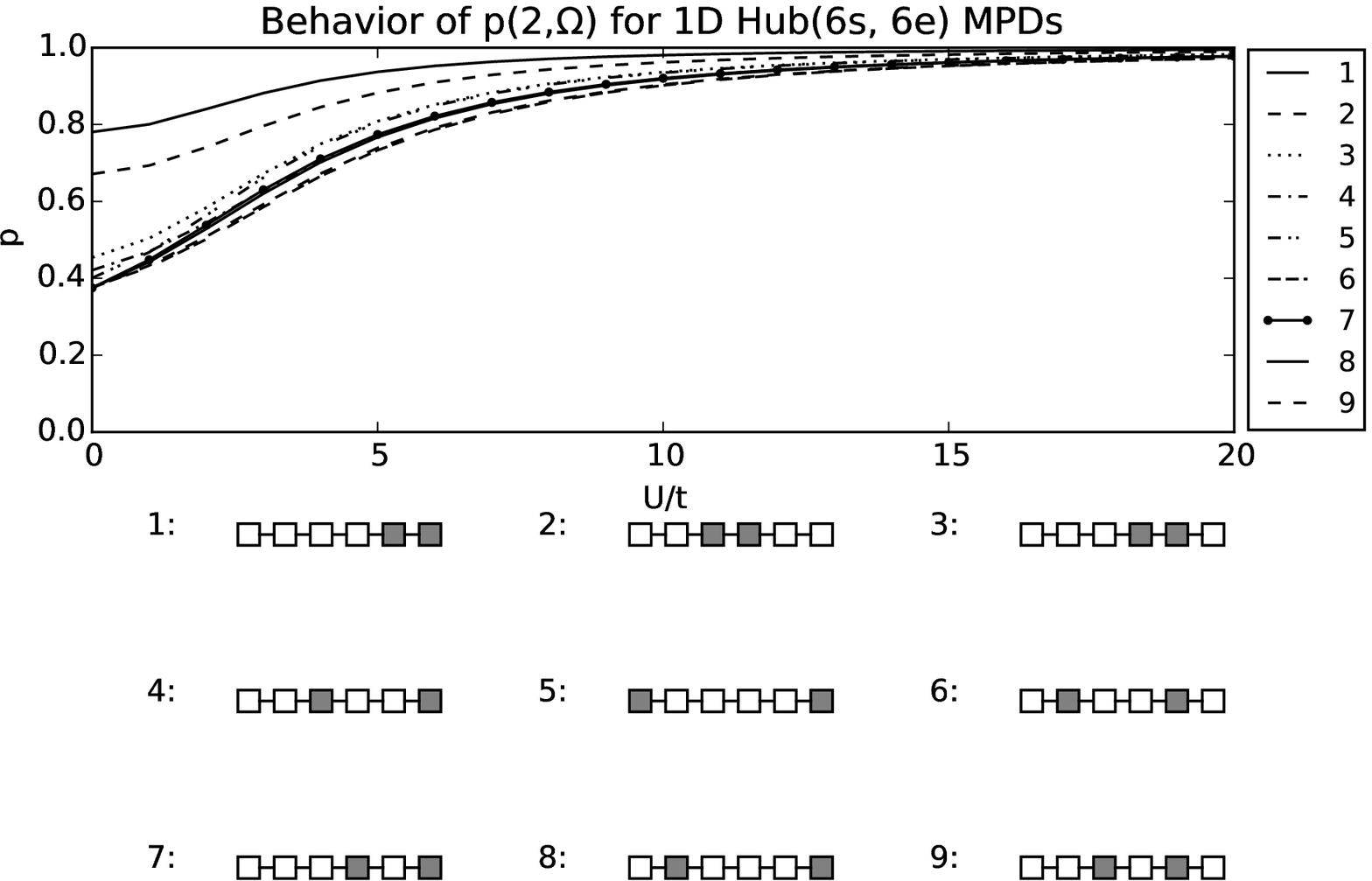}
    \caption{MPDs for choosing 2 spin-unresolved electrons for 1D Hubbard without periodic boundary conditions for 6 sites at half-filling for different $U/t$ values.}
    \label{fig:hub1_6s_6e_noper_mpd_2e}    
  \end{center}
\end{figure}

The 1D Hubbard model with 6 sites at half filling and no periodic boundary conditions can be used as a model for 1,3,5-hexatriene (Hubbard 1,3,5-hexatriene). If we determine the MPDs for 2 electrons (regardless of spin), we find the same 9 symmetry classes for all values in the $U/t$ range (Figure \ref{fig:hub1_6s_6e_noper_mpd_2e}).
\par
The MPDs consisting of terminal sites are more probable than the MPDs containing the interior sites, with a clear decrease in probability for other domain configurations. This is again in line with what we expect from the underlying physics: the on-site repulsion `pushes' the electrons towards the extremal sites. All probabilities again converge for high $U/t$ values due to the anti-ferromagnetic properties of the wave function.
\par
\begin{figure}[p]
  \begin{center}
    \includegraphics[width=\textwidth]{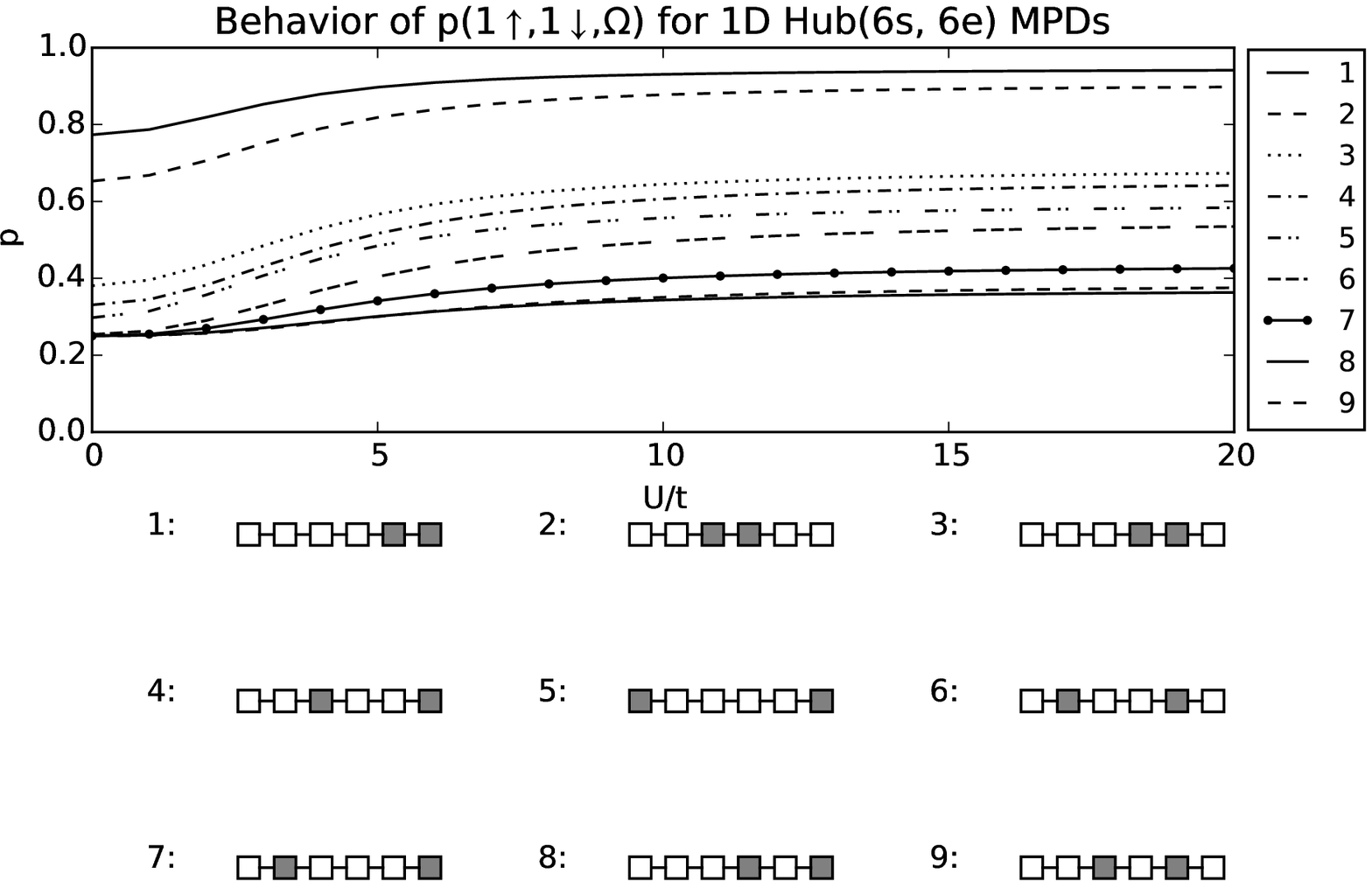}
    \caption{MPDs for choosing 1 up-spin and 1 down-spin electron for 1D Hubbard without periodic boundary conditions for 6 sites at half-filling for different $U/t$ values.}
    \label{fig:hub1_6s_6e_noper_mpd_1u_1d}
  \end{center}
\end{figure}
\begin{figure}[p]
  \begin{center}
    \includegraphics[width=\textwidth]{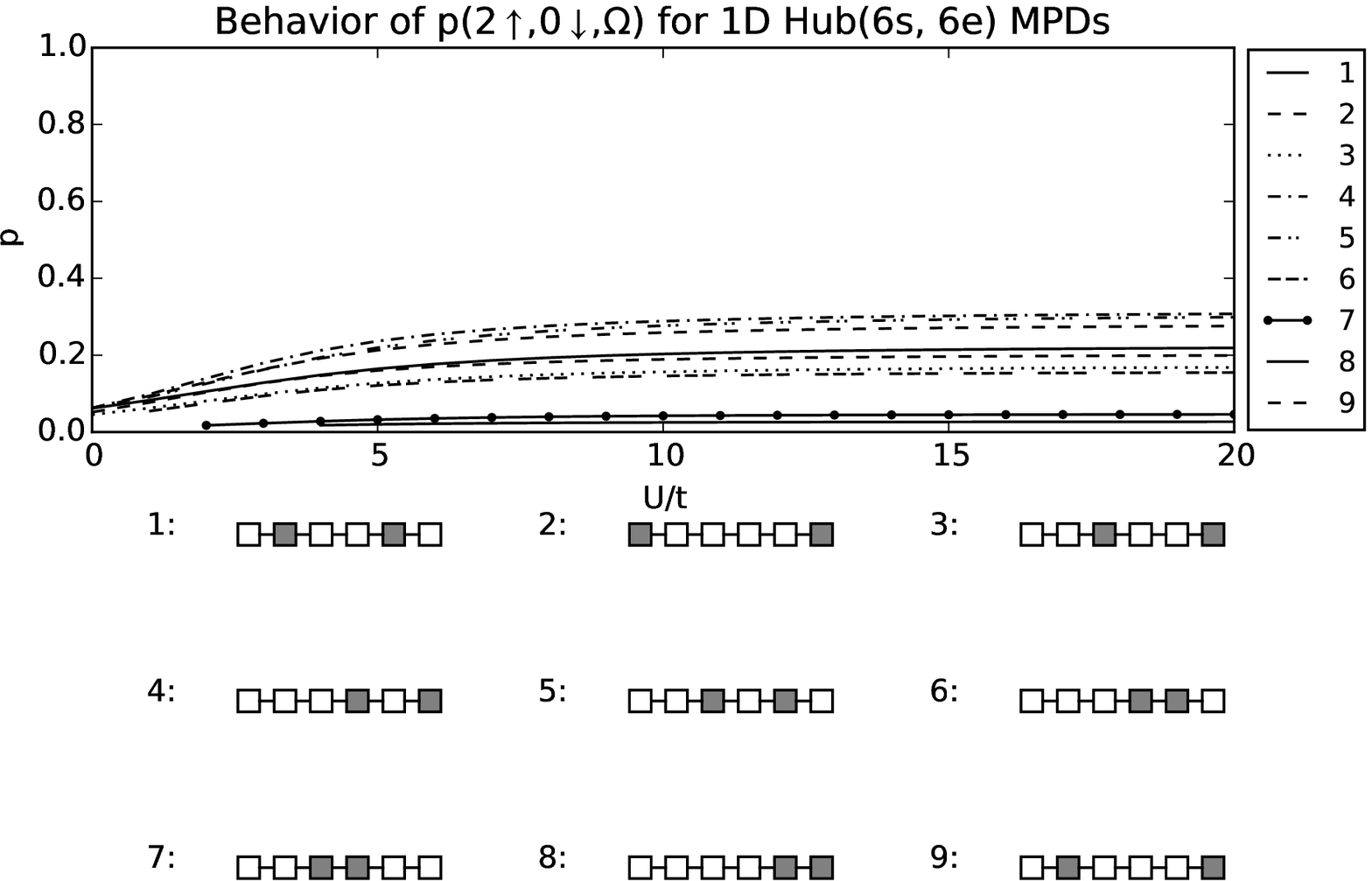}
    \caption{MPDs for choosing 2 up-spin and 0 down-spin electrons for 1D Hubbard without periodic boundary conditions for 6 sites at half-filling for different $U/t$ values.}
    \label{fig:hub1_6s_6e_noper_mpd_2u_0d}    
  \end{center}
\end{figure}

If we determine the spin-resolved MPDs (Figures \ref{fig:hub1_6s_6e_noper_mpd_1u_1d} and \ref{fig:hub1_6s_6e_noper_mpd_2u_0d}), we again obtain the same classes of MPDs as in the spin-unresolved case, although this is not dictated by the theory as such. For typical electron pairs, the highest probability is again found for MPDs consisting of two adjacent sites; the relative low probability of structure `3' for $\left\{ \mathrm{MPD}(1_{\uparrow}, 1_{\downarrow})\right\}$ can be linked to the required nodal structure of the underlying wave function. The probabilities associated to $\left\{ \mathrm{MPD}(2_{\uparrow}, 0_{\downarrow})\right\}$ are again lower compared to the two other choices.
\par
Again, the MPDs behave as can be expected for different correlation regimes. Furthermore, they appear to be in line with what is expected from Valence Bond Theory (see for instance the discussion of 1,3,5-hexatriene by \cite{wu2000a}).

\section{Concluding remarks and perspectives}

We have shown that the framework of the Hubbard model can be used to highlight the essential theoretical structure behind MPDs, both for FCI wave and HF wave functions. Furthermore, we have implemented a fast analysis procedure which can be used to analyse numerical results, and we have illustrated this with two applications at the FCI level. We have shown that the physics behind the considered Hubbard model can be captured using MPDs.
\par
It is important to note that the theory of MPDs allows the user to choose a number of electrons which is different than two for determining MPDs. For the small systems under scrutiny, these types of analyses typically provide less information. However, we are currently investigating larger systems, where we might be able to link MPDs to other chemical concepts.
\par
We also note that the problem of multi-domain optimisation, where several domains are optimised simultaneously for a given partition of electrons, becomes feasible in the Hubbard model using the method of projection operators. We are currently investigating algorithms along these lines.

\section*{Acknowledgements}

All authors are members of the QCMM alliance Ghent-Brussels and acknowledge financial support from the Research Foundation Flanders (FWO-Vlaanderen), G.A. and P.W.C. as pre-doctoral fellows. The authors also acknowledge the Special Research Fund of Ghent University for continuous support. The computational resources (Stevin Supercomputer Infrastructure) and services used in this work were provided by the VSC (Flemish Supercomputer Center), funded by Ghent University, the Hercules Foundation, and the Flemish Government - department EWI.

\bibliographystyle{unsrt}
\bibliography{hubbard_mpd}

\appendix

\section{Generating function}
\subsection{Factorisation of the generating function}\label{subsection:appendix:factorisation}
In the present appendix, we will make abstraction of the spin-$\sigma$ variable.  Before proving the factorisation of the generating function, it is instructive to consider a simple recurrence relation at the operator level.  Assume that we construct a domain $\Omega\cup \{i\}$ by adding a single site $i$ to an existing domain $\Omega$, then one has
\begin{equation}\label{appendix:recurrence1}
  \hat{P}(\nu,\Omega\cup\{i\})=\hat{P}(\nu,\Omega)\hat{P}(0,\{i\})+\hat{P}(\nu-1,\Omega)\hat{P}(1,\{i\}) \ .
\end{equation}
The proof goes via the explicit definition of the probability domain operator (\ref{theory:mpd:operator}).  Indeed, using the permutation symmetry among occupied ($O_{i}$) and unoccupied ($U_{i}$) operators, one can decompose the sum over all permutations into permutations restricted within the smaller domain and the single site $i$.  Note that for the special case of $\nu=0$\ , the recurrence relation simplifies to 
\begin{equation}
\hat{P}(0,\Omega\cup\{i\})=\hat{P}(0,\Omega)\hat{P}(0,\{i\}) \ .
\end{equation}

The recurrence relation (\ref{appendix:recurrence1}) has a nice classical interpretation.  The probability of finding exactly $\nu$ particles within an extended domain $\Omega\cup\{i\}$ is indeed the probability of finding exactly $\nu$ particles in $\Omega$ and zero particles at the site $i$, plus the probability of finding $\nu-1$ particles in $\Omega$ and the remaining particle at site $i$.  It is worth stressing though that this interpretation is strictly classical and is not necessarily true in general at the quantum level  
\begin{align}
P(\nu,\Omega\cup\{i\})&=\langle\Psi|[\hat{P}(\nu,\Omega)\hat{P}(0,\{i\})+\hat{P}(\nu-1,\Omega)\hat{P}(1,\{i\})]|\Psi\rangle,\\
&\neq\langle\Psi|\hat{P}(\nu,\Omega)|\Psi\rangle\langle\Psi|\hat{P}(0,\{i\})|\Psi\rangle\\
&\qquad+\langle\Psi|\hat{P}(\nu-1,\Omega)|\Psi\rangle\langle\Psi|\hat{P}(1,\{i\})|\Psi\rangle,\\
&=P(\nu,\Omega)P(0,\{i\})+P(\nu-1,\Omega)P(1,\{i\}) \ .
\end{align}
The equality only holds when the quantum state $|\Psi\rangle$ is an eigenstate of the projection operators (\ref{theory:projectionoperators}), which means that the resolution of the identity is equivalent to taking the expectation values of the projection operators.  This is explicitly the case for Slater determinant states in the site basis (\ref{theory:sitebasis}).

We are now at the position of proving the factorisation of the generating function at the operator level
\begin{equation}\label{appendix:generatingfunction:factorisation}
  \hat{G}(\Omega,t)=\sum_{\nu=0}^{M}\hat{P}(\nu,\Omega)t^{\nu}=\prod_{k=1}^M(\hat{U}_{i_k}+t\hat{O}_{i_k}) \ .
\end{equation}
The proof goes by induction.  Assume that the factorisation is valid for domains $\Omega$ of size $M$.  Adding one additional site $i$ gives rise to a domain $\Omega\cup\{i\}$ of size $M+1$.  The generating function for the latter domain is
\begin{align}
\hat{G}(\Omega\cup\{i\},t)&=\sum_{\nu=0}^{M+1}\hat{P}(\nu,\Omega\cup\{i\})t^\nu\\
&=\sum_{\nu=0}^{M+1}[\hat{P}(\nu,\Omega)\hat{P}(0,\{i\})+\hat{P}(\nu-1,\Omega)\hat{P}(1,\{i\})]t^\nu \ ,
\end{align}
in which we have used the recurrence relation (\ref{appendix:recurrence1}).  Careful reorganisation of both summation terms leads to  
\begin{align}
\hat{G}(\Omega\cup\{i\},t)&=\sum_{\nu=0}^M\hat{P}(\nu,\Omega)t^\nu\left(\hat{P}(0,\{i\})+t\hat{P}(1,\{i\})\right)\\
&=\hat{G}(\Omega,t)\left(\hat{U}_i+t\hat{O}_i\right) \ .
\end{align}
If we label site $i$ as $i_{M+1}$, the factorisation is proven.

It is possible to condense the generating function even further in notation using the Taylor expansion of the projection operator
\begin{equation}
e^{\alpha \hat{O}_i}=\sum_{k=0}^\infty \frac{\alpha^k}{k!}\hat{O}_i^k=1+\sum_{k=1}^\infty \frac{\alpha^k}{k!}\hat{O}_i=1+(e^\alpha-1)\hat{O}_i=\hat{U}_i+e^\alpha \hat{O}_i \ .
\end{equation}
Identifying $e^\alpha=t$, we can rewrite the generating function (\ref{appendix:generatingfunction:factorisation}) in exponential form
\begin{equation}
\hat{G}(\Omega,t)=\prod_{k=1}^M t^{\hat{O}_{i_k}}=t^{\sum_{k=1}^M\hat{O}_{i_k}}=t^{\hat{n}(\Omega)} \ ,
\end{equation}
with $\hat{n}(\Omega)$ the number operator counting the number of particles contained within the domain $\Omega$.
\subsection{Generating function for Slater determinants}\label{subsection:appendix:slater}
We will again make abstraction of the spin-$\sigma$ variable and consider a Slater determinant wave function with $N$ occupied orthonormal single-particle states
\begin{equation}
|\Phi \rangle = \prod_{\alpha=1}^N \left(\sum_{i=1}^L q_{i \alpha} \hat{a}_i^{\dagger}\right) |\theta\rangle \ ,
\end{equation}
in which $q_{i\alpha}$ denotes the unitary transformation from the site basis to the Slater determinant basis, with $\sum_{i=1}^L q_{i\alpha}^{*}q_{i \beta}=\delta_{\alpha \beta}$. It is now possible to evaluate the generating function 
\begin{align}
\langle \Phi |t^{\hat{n}(\Omega)} | \Phi \rangle &=\sum_{\substack{j_1, \dots j_N \\ k_1, \dots k_N}}^L q_{j_1 \alpha_1}^* \dots q_{j_N \alpha_N}^* q_{k_1\alpha_1}\dots q_{k_N \alpha_N} \nonumber \\
 &\qquad \qquad \times \langle \theta|a_{j_N} \dots \hat{a}_{j_1} t^{\hat{n}(\Omega)} \hat{a}_{k_1}^{\dag} \dots \hat{a}_{k_N}^{\dag} |\theta\rangle \\
 &=\sum_{j_1 \dots j_N}^L \sum_{\pi \in S_N}(-1)^{\pi}q_{j_1 \alpha_1}^* \dots q_{j_N \alpha_N}^* q_{\pi(j_1)\alpha_1}\dots q_{\pi(j_N) \alpha_N} t^{n_{\Omega}(j_1 \dots j_N)} \ ,
\end{align}
with $S_N$ the group of permutations of $[j_1 \dots j_N]$, $n_{\Omega}(j_1 \dots j_N)$ denoting the number of sites of $\Omega$ in $[j_1 \dots j_N]$, and where we have used the fact that $t^{\hat{n}(\Omega)}$ is diagonal in the site basis, with the fermionic anticommutation relations resulting in a phase factor. This can be further simplified to
\begin{align}
\langle \Phi |t^{\hat{n}(\Omega)} | \Phi \rangle &=\sum_{j_1 \dots j_N}^L\sum_{\pi \in S_N}(-1)^{\pi}q_{j_1 \alpha_1}^* \dots q_{j_N \alpha_N}^* q_{j_1 \pi(\alpha_1)}\dots q_{j_N \pi(\alpha_N)} t^{n_{\Omega}(j_1 \dots j_N)} \\
  &=\sum_{j_1 \dots j_N}^L \det\left[Q(j_1 \dots j_N)\right]t^{n_{\Omega}(j_1 \dots j_N)} \ ,
\end{align}
with $Q(j_1 \dots j_N)$ an $N \times N$ matrix with matrix elements $Q(j_1 \dots j_N)_{ik}=q^{*}_{j_i \alpha_i} q_{j_i \alpha_k}$. This can be seen as the expansion of a single determinant, where each column contains a summation over $j_i$ and is given a weight $t$ if $j_i \in \Omega$ and a weight 1 if $j_i \notin \Omega$. This results in
\begin{align}
\langle \Phi |t^{\hat{n}(\Omega)} | \Phi \rangle &=\det \left\{S(\bar{\Omega})+tS(\Omega) \right\} \ ,
\end{align}
where we have introduced two $N \times N$ matrices defined as
\begin{equation}
S(\bar{\Omega})_{jk}=\sum_{i \notin \Omega}q_{i \alpha_j}^{*}q_{i\alpha_k} \ , \qquad S(\Omega)_{jk}=\sum_{i \in \Omega}q_{i \alpha_j}^{*}q_{i\alpha_k} \ .
\end{equation}
The orthonormality condition results in $S(\bar{\Omega})+S(\Omega)=I_N$\ , which allows us to rewrite 
\begin{equation}
\langle \Phi |t^{\hat{n}(\Omega)} | \Phi \rangle =\det \left\{ \left[ I_N-S(\Omega) \right]+t S(\Omega) \right\} \ .
\end{equation}
This is the first determinant expression to be proven, which can be further rewritten by noting that $S(\Omega)=q(\Omega)^{\dag}q(\Omega)$\ , with $q(\Omega)$ an $M \times N$ matrix defined as $q(\Omega)_{i\alpha}=q_{i\alpha}$\ , with $i \in \Omega$. From Sylvester's determinant theorem, we have
\begin{equation}
\det \left\{ I_N+ (t-1) q(\Omega)^{\dag}q(\Omega) \right\}=\det \left\{ I_M+ (t-1) q(\Omega)q(\Omega)^{\dag} \right\} \ ,
\end{equation}
resulting in the second determinant expression for the generating function.

\end{document}